# Avalanche-Induced Current Enhancement in Semiconducting Carbon Nanotubes

Albert Liao[1], Yang Zhao[1], Eric Pop[1,2,*]

[1]*Dept. of Electrical and Computer Engineering, Micro and Nanotechnology Laboratory, University of Illinois, Urbana-Champaign, IL*

[2]*Beckman Institute, University of Illinois, Urbana-Champaign, IL*

Semiconducting carbon nanotubes under high electric field stress (~10 V/μm) display a striking, exponential current increase due to avalanche generation of free electrons and holes. Unlike in other materials, the avalanche process in such 1D quantum wires involves access to the third sub-band, is insensitive to temperature, but strongly dependent on diameter $\sim\exp(-1/d^2)$. Comparison with a theoretical model yields a novel approach to obtain the inelastic optical phonon emission length, $\lambda_{OP,ems} \approx 15d$ nm. The combined results underscore the importance of multi-band transport in 1D molecular wires.

[*]Contact: epop@illinois.edu
(please note the recent domain name change from uiuc.edu to illinois.edu)

Electrical transport in one-dimensional (1D) nanomaterials is of much fundamental and practical interest. Among these, single-wall carbon nanotubes have remarkably high performance, displaying quasi-ballistic transport at sub-micron dimensions [1], and excellent low-field mobility even in longer, diffusive samples [2]. Despite such progress, less is known about diffusive transport at high fields (>1 V/µm). This regime sets the peak current-carrying ability, and provides a glimpse into the behavior under extreme electrical stress conditions. For instance, the maximum current capacity of *metallic* single-wall nanotubes (m-SWNTs) is 20–25 µA when limited by Joule heating and optical phonon back-scattering [3, 4]. The maximum current capacity of *semiconducting* single-wall nanotubes (s-SWNTs) under diffusive transport is less established, although a 25 µA limit has also been suggested for single-band conduction [5]. However, experimental data indicates this limit is exceeded under ambipolar transport [6], and theoretical estimates also suggest this value can be surpassed when multiple sub-bands are efficiently involved [7].

In this Letter we report a remarkable exponential current increase beyond the 25 µA "diffusive limit" in Ohmically contacted s-SWNTs under avalanche impact ionization conditions. We investigate transport up to electrical breakdown and find the current in s-SWNTs first plateaus near ~25 µA, then sharply increases at high fields (~10 V/µm). This behavior is not seen in the many m-SWNTs tested. The current "up-kick" is attributed to the onset of avalanche impact ionization (II), a phenomenon observed in semiconductor p-n diodes and transistors at high fields [8-11], but not previously measured in nanotubes. We explore the behavior of s-SWNTs in such extreme conditions and demonstrate a novel approach for obtaining the optical phonon (OP) scattering length, which is the strongest energy relaxation process at high fields, and itself a fundamental transport parameter.

Carbon nanotubes were grown by chemical vapor deposition (CVD) from patterned Fe catalyst on 100 nm thermal $SiO_2$ and highly p-doped Si wafers which also serve as back-gates. The nanotubes were top-contacted by evaporating 40 nm of Pd, as shown in Figs. 1A and 1B. Electrode separation varied from $L$ ~ 1-4 µm and typical contact resistance was Ohmic, in the range $R_C$ ~ 30-50 kΩ, as judged from low-field *I-V* measurements. Metallic and semiconducting nanotubes were sorted by their on/off ratios, measuring current ($I_D$) vs. gate-source voltage ($V_{GS}$), as in Fig. 1C. As-grown devices show p-type behavior with negative threshold voltage ($V_T$). Dimensions were obtained by atomic force microscopy (AFM), indicating diameters were in the range $d \sim$ 2-3.6 nm.

Current vs. drain-source voltage ($V_{DS}$) measurements were made in air and vacuum. In air, metallic nanotubes saturate from self-heating and strong electron-phonon scattering [4] up to Joule breakdown, as in Fig. 1D. By contrast, most semiconducting tubes turned on at large $|V_{GS}|$ exhibit a sudden current increase before Joule breakdown (comparison in Fig. 1D). Additional measurements carried out in vacuum ($\sim 10^{-5}$ Torr, Figs. 2-5) allow further study of the current up-kick without breaking the nanotubes by oxidation. It is important to note that devices were measured in the reverse bias regime, with $V_{GS} < 0 < V_{DS}$ and $|V_{GS}| > |V_{DS}|$ [7]. By contrast, in Schottky mid-gap contacted devices, the ambipolar regime $V_{DS} < V_{GS} < 0$ "splits" the potential drop along the nanotube resulting in lower longitudinal electric fields [6, 7, 12] and transport by both electrons and holes. In the reverse bias regime, holes are the majority carriers in our s-SWNTs until the avalanche mechanism partially turns on the conduction band (Fig. 2A).

At first glance, several mechanisms may be responsible for the current increase at very high fields ($\sim 10$ V/μm) in our s-SWNTs, all various forms of "soft" (reversible) breakdown [14]. These are Zener band-to-band (BB) tunneling [13], avalanche impact ionization (II), and thermal generation current. Under BB transport electrons tunnel from the valence to the conduction band. The probability is evaluated as $P_{BB} \sim \exp(-E_G^2/q\hbar v_F F)$, where $E_G$ is the band gap ($\sim 0.84/d$ eV/nm), $v_F$ is the Fermi velocity, and $F$ is the electric field [13]. During avalanche II holes gain high energy in the valence band, then lose it by creating electron-hole pairs (EHPs) as shown in Fig. 2A. Inelastic optical phonon (OP) emission is the strongest competing factor to II, given the large OP energy ($\hbar\omega_{OP} \sim 0.18$ eV). The II probability is estimated as $P_{II} \sim \exp(-E_{TH}/q\lambda_{OP,ems}F)$ [8, 15, 16]. Here we take $\lambda_{OP,ems} \sim 14d$ nm as the spontaneous OP emission mean free path (MFP) by holes or electrons [16], and $E_{TH}$ is the avalanche energy threshold. Comparing the two mechanisms in Fig. 2B suggests impact ionization is considerably more likely for the electric field and nanotube diameter range in this study. BB transport does become important when bands are significantly more bent as a result of sudden spatial changes in electrostatic or chemical doping, leading to local fields of the order 100 V/µm (1 MV/cm) or higher [17, 18].

Previous theoretical work has shown II in s-SWNTs is not possible until the third sub-band is occupied [16], due to angular momentum conservation. Hence, the II threshold energy measured from the edge of the first band scales as $E_{TH} \sim 3/2E_G \sim 1.26/d$ (nm), which is greater than the band gap, as is typical in other semiconductors [10, 11]. To determine if the third sub-band is populated in our experiments, we look at the nanotube density of states (DOS), in Fig. 2C. Each Van Hove singularity represents the beginning of a sub-band. As the gate bias is lowered beyond threshold, the Fermi level inside the nanotube shifts to the right on the DOS plot and the third sub-band begins to fill at approximately $|V_{GS}-V_T| \sim 15$ V. The observed $V_T$ for our devices is in the range of -7 to -15 V. Thus, filling the third sub-band is easily within reach experimentally, also demonstrated in Fig. 3. In addition, we find direct injection into higher sub-bands at the contacts is also possible, as previously suggested by Ref. [7]. We estimate this in Fig. 2D using a Landauer-WKB integral to calculate the conductance associated with direct injection into the first three sub-bands at the Pd electrode. Naturally, injection into higher sub-bands depends strongly on voltage, and while direct injection into the third band is possible, we expect that high-field inter-valley scattering [19, 20] and gate-controlled charge distribution (Fig. 2C) are primarily responsible for populating the higher sub-bands.

The effects of gate voltage on the impact ionization tail for nanotube devices of different lengths, but similar diameters are shown in Fig. 3. First, for a given length, the same current up-kick is observed despite the value of $V_{GS}$, i.e. the four data curves converge on the "tail" region. Second, given the similar diameter (similar II threshold $E_{TH}$) it is apparent that the onset of the avalanche up-kick happens around the same approximate *field* (~$V_{DS}/L$), not the same drain voltage. The experimental results are in good agreement with our sub-band injection and DOS calculations (Fig. 2) where the occupation is dependent on $V_{GS}$ and the sub-band current contribution is dependent on $V_{DS}$.

An important characteristic of the avalanche process in many semiconductors such as silicon is the negative temperature dependence of the II coefficient [9]. As the phonon scattering rate increases with temperature, the free carriers gain energy less efficiently from the field, and the II rate decreases at higher temperatures. Here, such trends are examined in Fig. 4A, showing experimental data taken from 150 K to 300 K. Unlike in silicon, we observe negligible temperature dependence of the high-bias impact ionization tail region. The essential difference lies in that the optical phonon (OP) emission MFP ($\lambda_{OP,ems}$) varies minimally with temperature in SWNTs. As the OP energy is much greater than in other materials, the OP occupation $N_{OP} = 1/[\exp(\hbar\omega_{OP}/k_BT)-1]$ is very small, $\ll 1$, where $k_B$ is the Boltzmann constant. Following [4], the spontaneous OP emission MFP can be written as $\lambda_{OP,ems} = [N_{OP}(300)+1]/[N_{OP}(T)+1]\lambda_{OP,300}$ where $\lambda_{OP,300} \approx 14d$ [16]. This MFP is shown for two diameters as the inset to Fig. 4A, illustrating the negligible temperature variation. The lack of temperature dependence and that of a significant current (Joule heating) dependence of the up-kick also indicates there is no significant contribution from thermal current generation. Quite the opposite, given the generation of EHPs rather than OPs during II, a lowered Joule heating rate in the highest field region near the drain is expected.

In order to better understand the field dependence of II, we have modified an existing SWNT model [4] by including II as an additional current path through a parallel resistor. The choice is motivated by the physical picture in Fig. 2A, which shows electron transport in the conduction band "turning on" as an additional channel at fields high enough to induce hole-driven II. The expression for this resistor is given as $R_{II} = R\exp(E_{TH}/q\lambda_{OP,ems}F)$, where $R$ is for single-band transport, computed self-consistently with the SWNT temperature. The results are shown in Fig. 4B with $\lambda_{OP,ems}$ included as mentioned above, and without any other adjustable parameters. Despite being an "augmented" single-band model, the simulation correctly captures the experimentally observed current up-kick and its delayed voltage onset (also see Fig. 1D). The straightforward analysis also allows us to gain physical insight into the avalanche process, and to intuitively extract a few more key parameters.

In the parallel resistor approach, the avalanche current is $I_{II} \approx I_S \exp(-E_{TH}/q\lambda_{OP,ems}F)$, where $I_S$ is the saturation current reached before II becomes significant. Inserting the expected diameter dependence $E_{TH} \approx E_1/d$ and $\lambda_{OP,ems} \approx \lambda_1 d$, we obtain $I_{II} \approx I_S \exp(-E_1/q\lambda_1 F d^2)$, where $E_1$ and $\lambda_1$ are the threshold energy and MFP for a nanotube of diameter 1 nm. Consequently, the average field at which $I_{II} = I_S/2$ is given by $<F_{TH}> \approx E_1/q\lambda_1 d^2 \ln(2)$. We empirically extract this field in Fig. 5A and plot it vs. $1/d^2$ for several diameters ($d \sim 2.2$-$3.6$ nm) in Fig. 5B. The slope of the linear fit thus scales as the ratio between the II threshold energy and the inelastic MFP, $E_1/\lambda_1$. However, the avalanche process is a strong function of the field, and most EHPs are generated at the peak, $F_{TH,MAX}$. The latter is estimated by noting that the potential near the drain has a dependence $V(x) \approx \ell F_0 \sinh(x/\ell)$, where $F_0 \sim 1$ V/μm is the saturation velocity field and $\ell$ is an electrostatic length scale comparable to $t_{ox}$ [21, 22]. Fitting this expression to our voltage conditions and nanotube dimensions, we find $F_{TH,MAX}/<F_{TH}> \sim 4.5$ for $L \sim 1$ μm device, and 3.5 for $L \sim 2$ μm. Thus, using the peak instead of the average field, the empirically extracted slope gives $E_1/\lambda_1 \sim 0.088$ eV·nm, where we take $E_1 = 1.26$ eV as the bottom of the third sub-band. Accounting for fit errors, this yields $\lambda_1 = 15 \pm 3$ nm as the inelastic OP emission MFP for $d = 1$ nm, or generally $\lambda_{OP,ems} = \lambda_1 d$. This value is in good agreement with the theoretically predicted $14d$ nm in Ref. [16], and our approach demonstrates an additional, novel empirical method for extracting this important transport parameter from high-field electrical measurements.

Before concluding, it is interesting to compare our results to those of Marty *et al* [23], who studied exciton formation during high-field unipolar transport in SWNTs. They observed radiative exciton recombination at high fields, but did not observe the dramatic current increase before breakdown. This was reasonably attributed to direct exciton annihilation, rather than the avalanche generation of free carriers. By contrast, our nanotubes have ~2x larger diameters and correspondingly smaller band separations and exciton binding energies, and Ohmic Pd contacts rather than highly resistive Co contacts. In addition, all our measurements except Fig. 1D were carried out in vacuum, allowing repeated study of the current up-kick which was not always observable in air before Joule breakdown. While excitonic generation and recombination may play a role in our samples, we suggest that the current increase is possible because most free EHPs are generated in the high-field region within a few mean free paths (10-100 nm) of the drain. Thus, generated electrons are swept out into the electrode by the high field within 0.1-1 ps (Fig. 2A), much faster than the recombination lifetimes (10-100 ps) [24]. In addition, the high temperatures and high fields in these conditions will contribute significantly to exciton instability, despite their relatively high binding energy.

In summary, we observe a remarkable current increase beyond 25 µA in semiconducting SWNTs driven into avalanche impact ionization at high fields (~10 V/µm). By analyzing near-breakdown *I-V* data, we find the avalanche process to be nearly temperature independent, but strongly diameter dependent $\sim\exp(-1/d^2)$, unlike in other materials. In addition, a novel estimate of the inelastic optical phonon scattering length $\lambda_{OP,ems} \approx 15d$ nm is obtained by fitting against a model of the high-field current "tail." We note that upper sub-band transport must be considered at high bias, and has a significant effect on the current carrying capacity of such nanomaterials. The results also suggest that avalanche-driven devices with highly non-linear characteristics can be fashioned from semiconducting carbon nanotubes.

We acknowledge valuable discussions with D. Jena, M. Kuroda, J.-P. Leburton and M. Shim. This work was supported in part by the Arnold O. Beckman award from the University of Illinois, by the NSF Network for Computational Nanotechnology (NCN), and by the Nanoelectronics Research Initiative (NRI) through the Midwest Institute for Nanoelectronics Discovery (MIND).

**Figure Captions**

**Figure 1:** Metallic and semiconducting SWNTs up to breakdown in air. (A) Schematic cross-section of back-gated nanotube. (B) Scanning electron microscope (SEM) top-view image of a fabricated device. Semi-circular electrodes are used for tighter control of device length. Scale bar is 10 µm. (C) Back-gate voltage dependence ($V_{GS}$) of semiconducting and metallic SWNT showing typical on/off ratios. (D) Drain voltage ($V_{DS}$) dependence up to breakdown in air of semiconducting and metallic SWNTs. Metallic device saturates before breakdown, whereas semiconducting tube displays an up-kick in current caused by avalanche impact ionization. Compared devices have similar diameter $d \sim 2.5$ nm and length $L \sim 0.8\text{-}1.1$ µm.

**Figure 2:** Theoretical basis for avalanche behavior of s-SWNT. (A) Schematic band diagram and EHP generation under reverse bias conditions. (B) Probability of impact ionization (II) and Zener band to band tunneling (BB) vs. electric field along the nanotube, for the diameters and field range of interest. (C) Computed density of states (DOS) showing the first four sub-bands. The second band begins to fill at $|V_{GS}\text{-}V_T| \sim 5$ V and the third at $|V_{GS}\text{-}V_T| \sim 15$ V, as pictured. (D) Contact conductance of the first three sub-bands under direct injection from the Pd electrode. The arrow indicates approximate voltage at which direct injection into the third sub-band becomes significant. Plots (C) and (D) obtained for $d = 2.5$ nm and $t_{ox} = 100$ nm.

**Figure 3:** Length dependence of impact ionization tail. Measured reverse bias current vs. drain voltage ($V_{DS}$) in vacuum with applied back-gate $V_{GS}$ for two s-SWNTs with similar diameter ($d \sim$ 2.5 nm) but with device lengths of (A) $L \sim 1.3$ µm and (B) $L \sim 2.3$ µm. The onset voltage for the avalanche "up-kick" scales as the lateral *field* and appears independent of $V_{GS}$.

**Figure 4:** (A) Temperature insensitivity of impact ionization tail. Measured reverse bias $I_D$-$V_{DS}$ curves for a s-SWNT with $d \sim 2.2$ nm and $L \sim 2.2$ µm, in vacuum. There is little temperature dependence of the avalanche behavior, unlike in other materials. This is attributed to the very small temperature dependence of the band gap and the optical phonon emission mean free path in nanotubes ($\lambda_{OP,ems}$ calculated for two diameters, following Ref. [4] in the inset). (B) Model including

and excluding impact ionization as a second parallel channel which begins to open up at high field (see text, also Figs. 1D and 2A).

**Figure 5:** Diameter dependence of avalanche threshold field, $F_{TH}$. (A) Current vs. average channel field $(V_{DS}-I_D R_C)/L$ for several s-SWNT diameters. The II threshold is extrapolated from the tail region and defined as the field $<F_{TH}>$ at which the current reaches one half the saturation value. (B) Extracted average $<F_{TH}>$ vs. $1/d^2$. The uncertainty in diameter from AFM measurements is 0.4 nm. The slope of the linear fit scales as the ratio between the avalanche energy threshold and the inelastic OP emission length, $E_{TH}/\lambda_{OP,ems}$. Taking $E_{TH} \sim 1.26/d$ eV, the OP emission length is $\sim 15d$ nm.

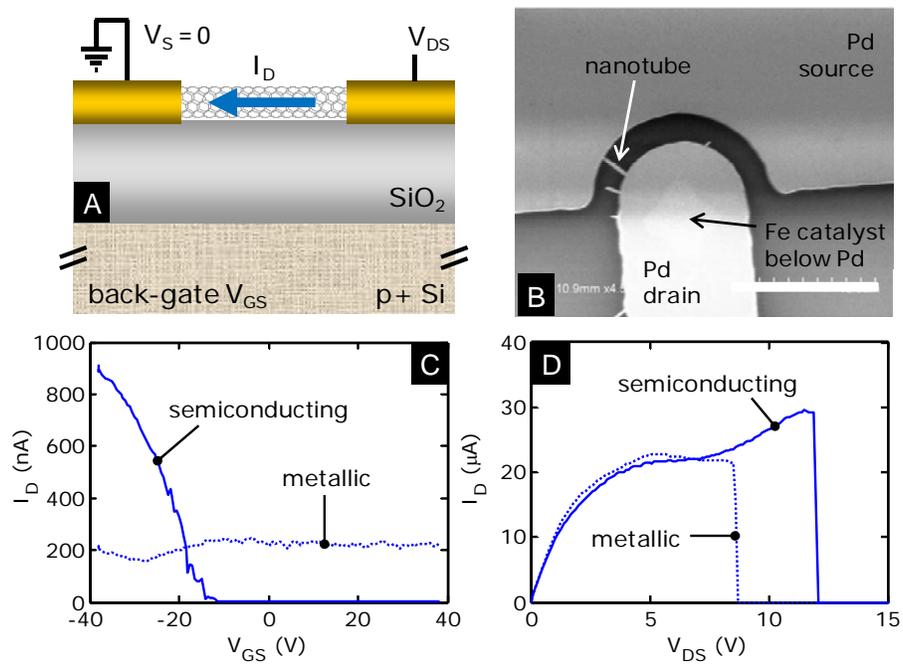

Figure 1.

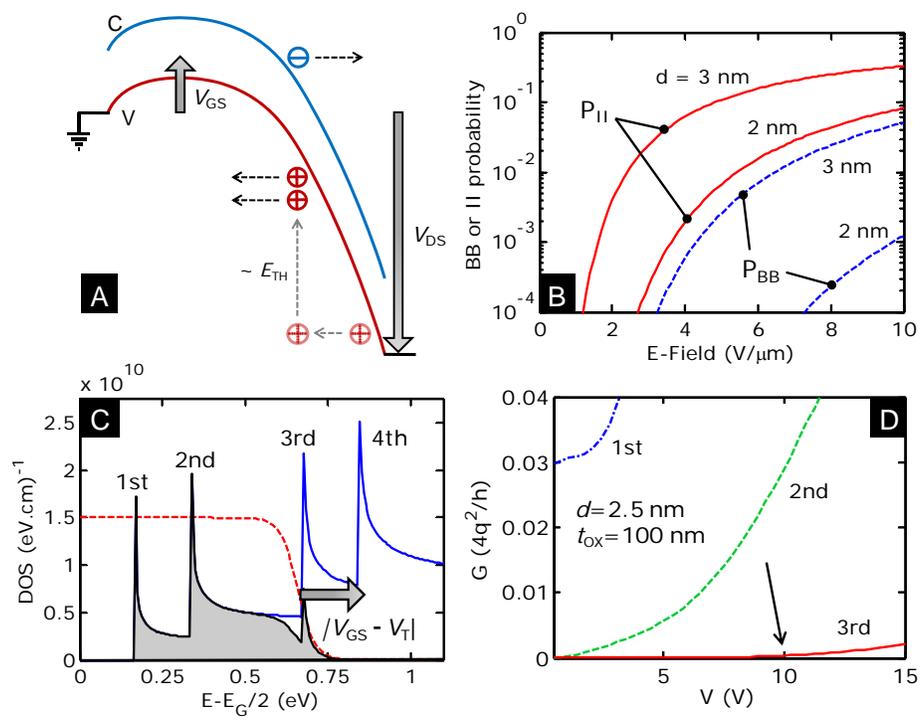

Figure 2.

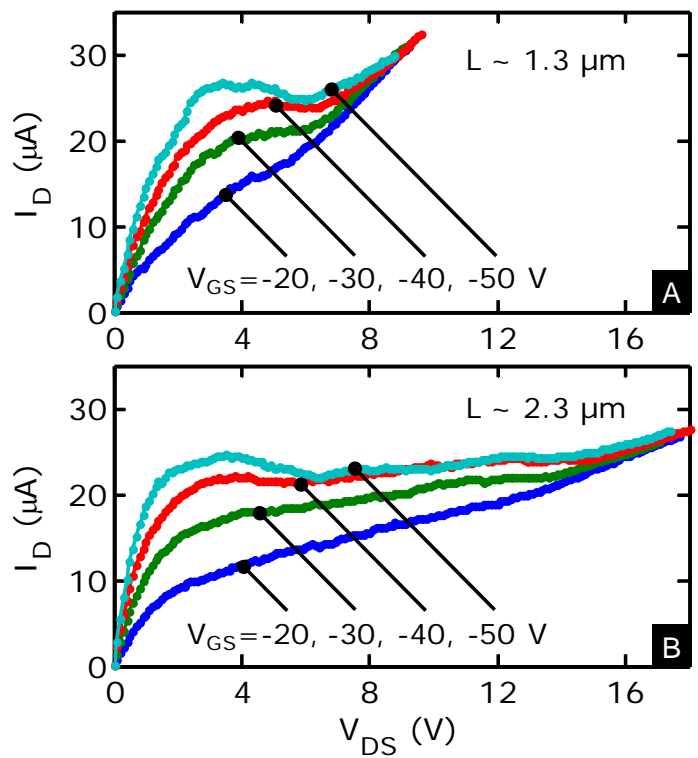

Figure 3.

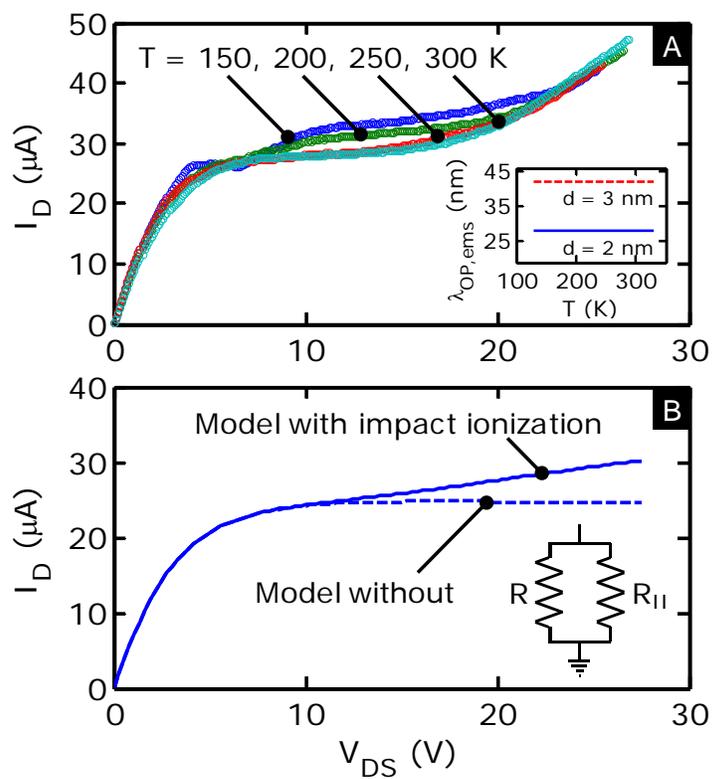

Figure 4.

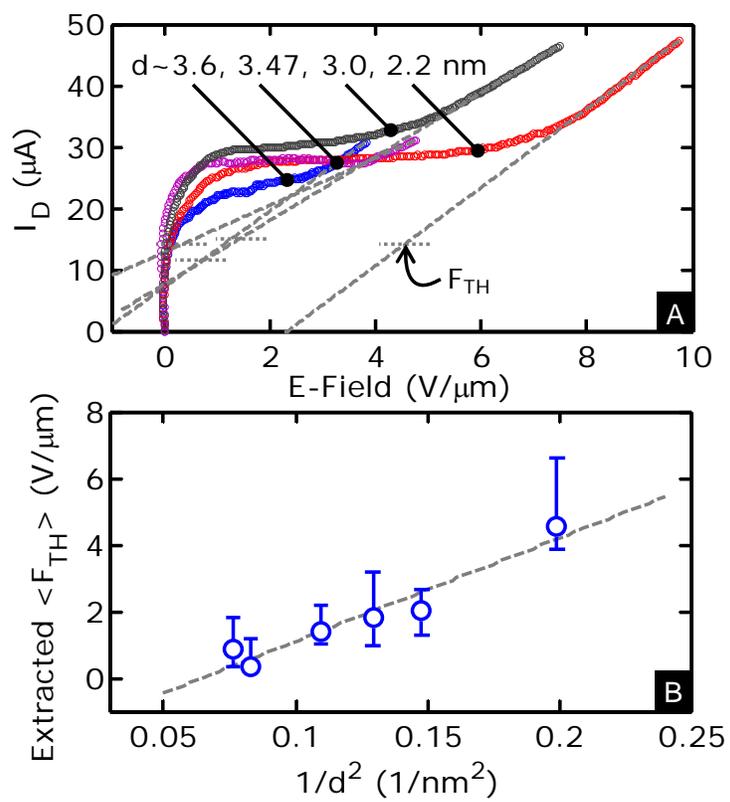

Figure 5.